\documentclass[aps,prb,preprint]{revtex4}
\usepackage{tabularx,graphicx}

\def\gdi{GdI$_2$\,\,} 
\def\gdx{GdI$_{2}$H$_{x}$\,\,} 

\begin{document}

\author {T. Maitra$^{1}$\footnote{email: maitra@itp.uni-frankfurt.de}, 
A. Taraphder$^{2,3}$, A. N. Yaresko$^{2}$ and P. Fulde$^{2}$}   
\title {Magnetic properties of doped \gdi}
\affiliation{ $^{1}$Institut f\"ur Theoretische Physik, J. W. Goethe 
Universit\"at, Max-von-Laue-Strasse 1,
 60438 Frankfurt am Main, Germany \\ 
$^{2}$ Max Planck Institut f\"ur Physik Komplexer Systeme, 
N\"othnitzer strasse 38, 01187 Dresden, Germany \\ 
$^{3}$Department of Physics and  Centre for 
Theoretical Studies,\\ Indian Institute of Technology, Kharagpur 721302 India }
\begin{abstract}
Motivated by the recent experimental studies on layered ferromagnetic 
metallic system \gdi and its doped variant \gdx we develop a model
to understand their ground state magnetic phase diagram. Based on first 
principle electronic structure calculations we 
write down a phenomenological model and solve it under certain approximations
to obtain the ground state energy. In the process we work out the phase 
diagram of the correlated double exchange model on a triangular lattice 
for the specific band structure at hand. 

\end{abstract}


\maketitle 
\vspace{.5cm} 

\section{Introduction}

Layered magnetic systems with coupled charge and spin degrees of
freedom have received considerable attention recently owing
to their rich phase diagram and intriguing transport properties. 
The layered system \gdi and its doped variant \gdx, recently studied 
by Felser et al.\cite{FAK+99} and Ryazanov et al.\cite{rya05}, show
significant magnetoresistance\cite{AFK+00} of about 60\% at room temperature
with a ferromagnetic (FM) transition temperature close to 300K\cite{kas84}.
Unlike the magnetoresistive manganites or their bilayer counterparts,
there is no significant Jahn-Teller coupling in these systems and 
that enables one to study a purely electronic model, coupled to
magnetic degrees of freedom, without the involvement of the lattice.  
In addition, \gdi is isostructural to the well-known dichalcogenides
like 2H-TaS$_2$,  2H-TaSe$_2$ with hexagonal layered structure, but 
do not show any superconducting instability - in fact the resistivity 
rises at low temperatures.

It has been pointed out recently\cite{ETF+01} that the large magnetoresistance
in \gdi is primarily due to the freezing out of the spin disorder
scattering of the conduction electrons in a magnetic field. There is 
indeed evidence for 
spin fluctuation coupled to charge degrees of freedom in the
ESR experiments\cite{dei}. The broad resistivity anomaly at Curie 
temperature (T$_c$) shifts towards higher temperatures with magnetic field.
From susceptibility measurements\cite{rya05} a variety of regions with 
different degree of spin ordering have been proposed. The high-temperature 
ferromagnetic phase gives way to regions of frozen or short-range correlated 
spins at lower temperatures. At about 33\% hole doping a sliver of spin 
glass phase appears at low temperatures ($T < 20$K) changing over to 
a paramagnetic (PM) phase with doping.  

In \gdi the rare earth Gd ions have one electron in the $d$-band coupled to
the localised Gd $4f$ electrons through ferromagnetic exchange in the
configuration $4f^{7}5d^{1}$. The rare earth ions are arranged in a 
hexagonally coordinated layer structure, each Gd ion has six nearest-neighbour
Gd ions. Each such layer is separated from the next one by two layers of
iodine atoms. This results in having the inter-layer coupling among the
Gd ions considerably weaker than the intra-layer one. LDA band structure 
results indicate a spin splitting of the conduction band \cite{FAK+99} 
- the splitting is nearly complete in the LDA+U results.  
The crystal field in the trigonal prismatic 
layered dichalcogenide is negligible and the $5d$ orbitals are quite strongly
mixed. Band structure calculations (see below) indicate that there is one
half-filled $d$-band that crosses the Fermi level. The on-site Coulomb 
interaction in the rare earth $5d$ level is generally not very strong.
But a single, half-filled narrow $d$-band crossing the Fermi level and 
low dimensionality of the system make this repulsion quite relevant. The 
presence of short-range spin fluctuations and spin disordered phases 
already point to competing magnetic exchange interactions in the system.
In the half-filled single $d$-band, even a small Coulomb repulsion 
generates AF fluctuations that will compete with the FM interaction
mediated by the conduction band via the FM $f-d$ exchange (similar
to the Zener double exchange (DE) mechanism). It is also useful to
note that the \gdi system never shows full saturation moment, predicted
by the LDA calculations. The ordered state has a moment of 7.33$\mu_B$,
less than the full saturation moment of 8$\mu_B$ out of which 7$\mu_B$
presumably comes from the $4f$ core spins\cite{FAK+99}. Hence the moment coming from 
the 5$d$ electron is considerably less than 1$\mu_B$ 
indicating that correlation effects (beyond the density functional 
calculations) and coupling of the spin and charge degrees of freedom
are very relevant. Also there is almost no literature on the study of a 
correlated DE model on a non-bipartite lattice. On a triangular lattice, 
it is known that the nearest-neighbour Ising AF interactions are frustrated 
leading to a finite ground state entropy. Indeed, the ground state of 
AF Heisenberg spin model or any correlated electronic model on a triangular 
lattice is not known. In the few studies that exist on frustrated itinerant 
systems \cite{lac+01}, the geometric frustration of the lattice has often 
been replaced by a random spin exchange. For a review on these and 
related issues, see, e.g., Ref. \onlinecite{Ful04}. 

\section {LSDA band structure}

Spin-polarized band structure calculations were performed for the
experimental crystal structure of GdI$_2$ \cite{FAK+99} within local
spin-density approximation (LSDA) using the LMTO method \cite{And75} in the
atomic sphere approximation with the combined correction term taken into
account. Gd 4$f$ electrons were assumed to be completely localized and
treated as quasi-core states. All seven majority spin 4$f$ states were
occupied while the minority spin ones remain empty which models the 
high-spin state of the Gd 4$f$ shell. Ferromagnetic arrangement of Gd 
magnetic moments was assumed. 

\begin{figure}[tbp!]
\centerline{\includegraphics[width=0.45\textwidth,clip]{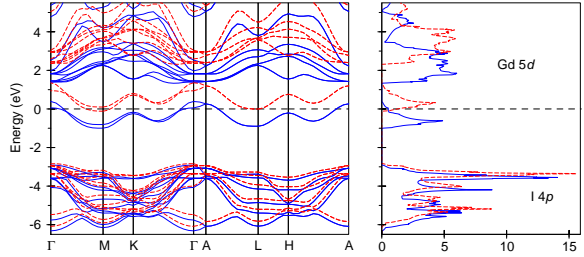}}
\caption{\label{fig:bnd}(Color online)
LSDA band structure and the total DOS of GdI$_2$. Majority and minority
spin bands are plotted with solid and dashed lines, respectively. }
\end{figure}

The calculated band structure shown in Fig.~\ref{fig:bnd} agrees well with
the results of previous calculations \cite{FAK+99}. I 5$p$ states are
completely filled and separated from partially occupied Gd 5$d$ states by a
gap of $\sim$2 eV. The latter are split into majority and minority spin
subbands by strong exchange interaction with the completely spin-polarized
4$f$ shell, the splitting being of order $\sim$0.9 eV. The total width of 5$d$
derived bands is 5 eV. The most striking feature of the band structure is
that for each spin direction two lowest Gd 5$d$ bands are split off from
the others by a gap of 1 eV. 

\begin{figure}[tbp!]
\centerline{\includegraphics[width=0.45\textwidth,clip]{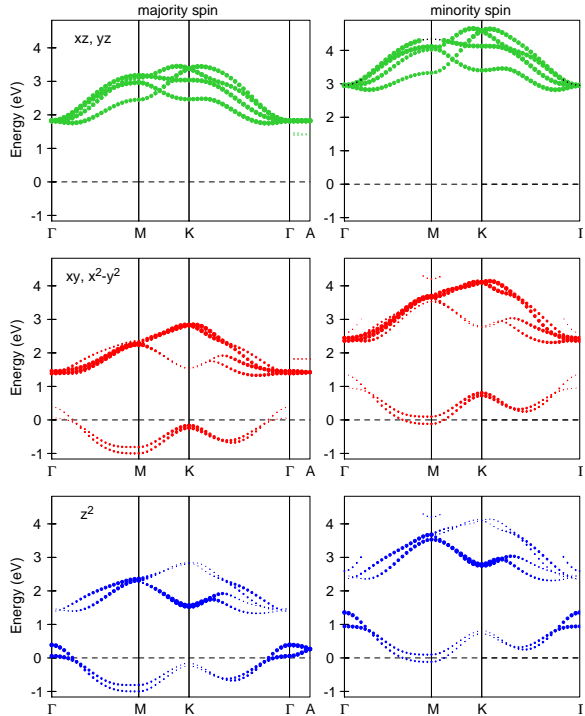}}
\caption{\label{fig:fb}(Color online) The expanded view of the Gd 5$d$
bands. The size of filled circles is proportional to the squared
contribution of 5$d$ orbitals of different symmetry to the Bloch wave
function at a given $k$ point. The coordinates of the high symmetry points
are $\Gamma$ (0,0,0), $M$ ($\frac{1}{\sqrt{3}}$,0,0), $K$
($\frac{1}{\sqrt{3}}$,$\frac{1}{3}$,0), and $A$ (0,0,0.136) in $2\pi/a$
units; $a$ being the lattice constant.}
\end{figure}

Because of the trigonal local symmetry of a Gd site $d$ orbitals are split
into a one-dimensional representation $a_1$ ($d_{z^2}$) and two
two-dimensional representations $e_1$ ($d_{xz}$, $d_{yz}$) and $e_2$
($d_{xy}$, $d_{x^2-y^2}$).
Band plots show the contribution of different Gd 5$d$
orbitals to a Bloch wave function represented by the size of a dot
(Fig.~\ref{fig:fb}).
One can glean that the assumption\cite{FAK+99} of two lowest Gd $d$ bands
formed predominantly by a $d_{z^2}$ orbital is not confirmed by our
calculation. 
Except for the $k$ space region around the $\Gamma-A$ high
symmetry direction, the contributions of $d_{z^2}$, $d_{xy}$, and
$d_{x^2-y^2}$ orbitals to the bands in question as well as to the two bands
above the gap are comparable. Thus, the gap appears as a result of a rather
strong hybridization between the above orbitals in the $ab$ plane. Four
highest Gd $d$ bands are formed by $d_{xz,yz}$ states which are shifted
towards higher energies due to the trigonal component of the ligand field.
They do not hybridize with the former three orbitals. The same conclusion on
the nature of the gap was drawn by Mattheiss \cite{MAT73} based on a tight
binding fit to the calculated band structure of a layered 2H-TaSe$_2$ compound
which is isostructural to GdI$_2$.

\begin{figure}[tbp!]
  \begin{center}
\includegraphics[width=7 cm]{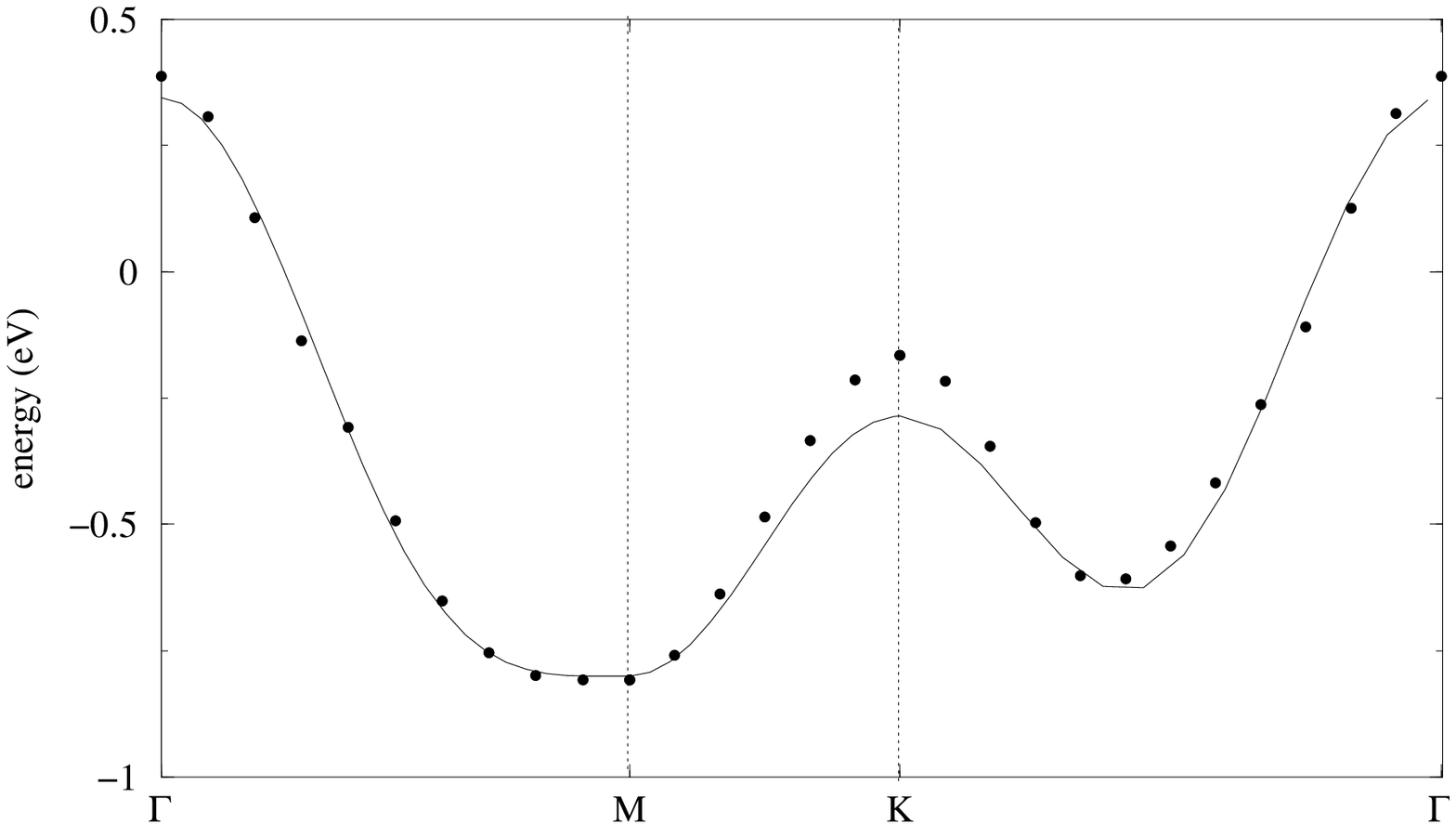}
\includegraphics[width=7 cm]{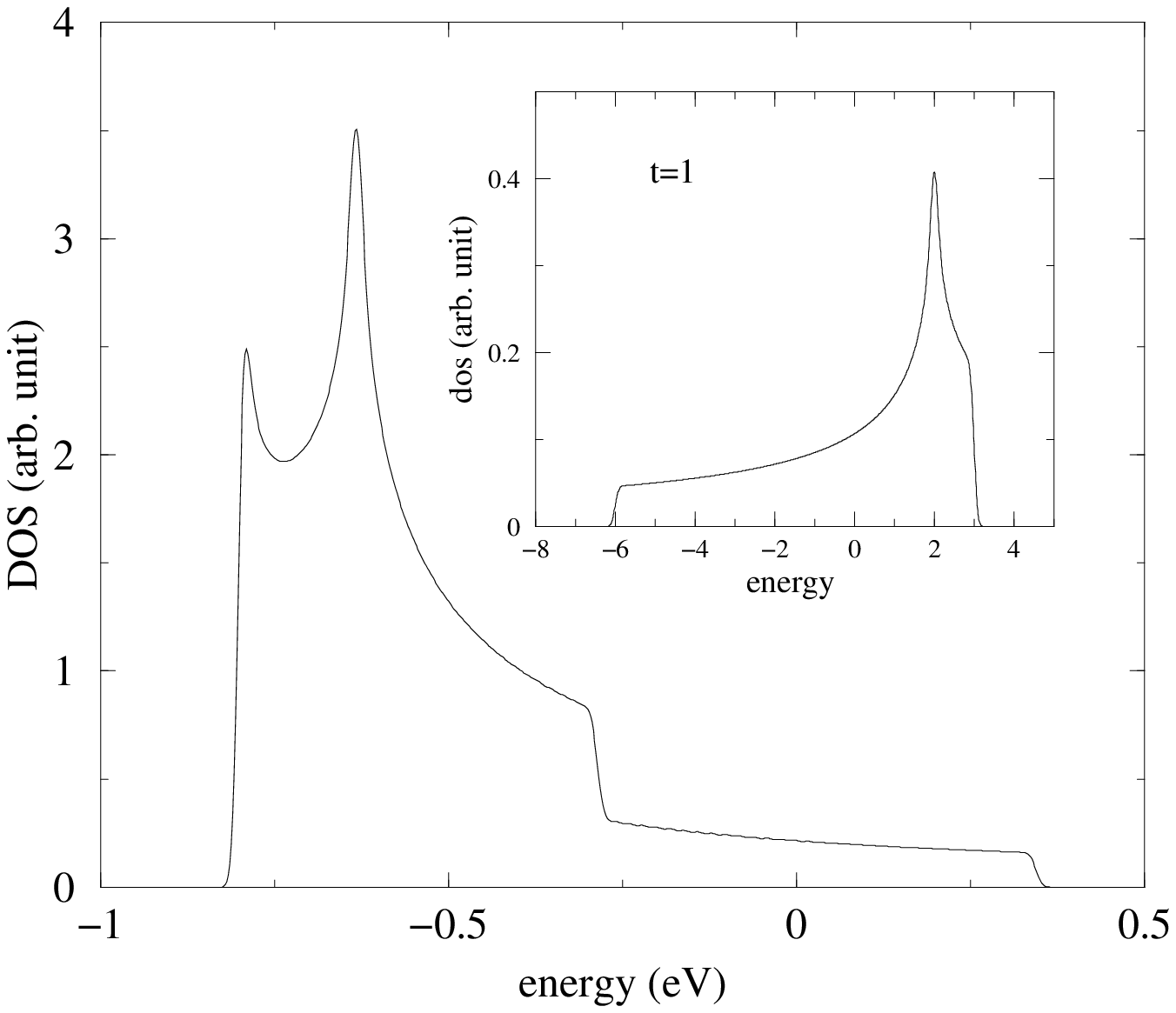}
\end{center}
    \caption{ (a) Tight binding fit (line) for the $5d$ band (dotted line) 
              that crosses the Fermi level in Fig. 1, (b) DOS corresponding
              to the tight binding band in (a) with the inset showing the 
              DOS for the tight binding band considered in the calculations of
              Hanisch et. al.\cite{mh95} (see text)} 
    \label{figure 3}
\end{figure}

\section{Effective Model}

It is clear from the band structure calculations that the band crossing the
Fermi level is a mixture of primarily three d-orbitals, namely d$_{z^2}$,
d$_{xy}$ and d$_{x^{2}-y^{2}}$. It is therefore not possible to obtain a
one-parameter, tight-binding fit of this band with only nearest neighbour
hopping\cite{footnote}. We fit this band using a three parameter ($t_0$,
$t_{nn}$ and $t_{nnn}$ = -0.514, 0.070 and 0.073 eV) tight binding one
(Fig. 3a). The band width is about 1eV and we assume the on-site repulsion
to be close to or less than this. In any case, we treat this as a parameter
in our calculation.  The corresponding density of states (DOS) is shown in
Fig. 3b along with the DOS of the one-parameter nearest neighbour DOS used
by Hanisch et al\cite{mh95} in their study of the Hubbard model on a
triangular lattice. The values of $f-d$ exchange have been estimated from
the above band structure results assuming a mean-field description for the
exchange and a classical description for the core spin. Assuming further
that the spin splitting is entirely due to $f-d$ exchange, the value
$J_{fd}S=0.9 eV$ then serves as an upper limit for the value of the
exchange coupling ($S=7/2$ is the $4f$ core spin).

Based on the above arguments, a possible Hamiltonian for the effective 
degrees of freedom can be written as 

\begin{widetext}

\begin{equation}
H = -J \sum_{<ij>} {\bf S_{i}}.{\bf S_{j}} - J_{fd} \sum_{i} {\bf S_{i}}.
{\bf s_{i}} - \sum_{<ij>\sigma}t_{ij} c_{i,\sigma}^{\dagger} c_{j,\sigma} 
+ \frac{U}{2} \sum_{i,\sigma} \hat{n}_{i\sigma} \hat{n}_{i-\sigma} 
\end{equation}
\end{widetext}

\noindent Here $J$ is the exchange interaction between the localised 4f
spins ${\bf S}_i$ ($S=7/2$). The second term is the $f-d$ FM exchange coupling
(${\bf S, s}$ are the core and conduction spin, respectively) while the 
third term represents the hopping of conduction electrons. 
The last term represents the Hubbard interaction.
In second-order perturbation theory, $J\sim -|t_{ij}|^2/(J_{fd}S+U)$ is  
due to the virtual exchange of  conduction electrons between
sites $i$ and $j$. It turns out to be antiferromagnetic (AF). Its magnitude
is small for the parameters discussed above. Away from half
filling, the DE mechanism favours an FM configuration of the core
spins driven by the kinetic energy (KE) of the electrons, while the
Coulomb repulsion $U$ favours an antiferromagnetic arrangement. 
This leads to competing ground states tuned by filling, with additional
richness coming from the intrinsic geometric frustration of an AF state 
defined on a triangular lattice. In the following, we keep $J=0$ and 
study the magnetic phase diagram of the model. We treat core spins
(S=7/2) semi-classically and use mean-field theory for the Coulomb
repulsion $U$.  

The effective 5d band at the Fermi surface in the undoped system ($x=0$) is
half-filled i.e., the electron density $n=1$. Hole (electron) doping,
therefore, implies $n < 1 \,\,(n > 1)$. Doping of holes in the \gdi is
effected by the insertion of hydrogen \cite{rya05} which localises
electrons away from the band. This is taken care by suitably moving the
Fermi energy. In the limit of $J_{fd}=0$ and with a nearest-neighbour
tight-binding band, this model represents the single band Hubbard model on
a triangular lattice studied within mean-field theory by several groups
\cite{mh95,hrk90,hrk91,gaz94,gaz97,machida92}.  These studies find a number
of magnetically ordered phases, i.e., the FM state, the Ne\`el state
[ferromagnetic rows coupled antiferromagnetically, the ordering wave vector
${\bf Q}=(2\pi/{\sqrt 3},0)$], the AFM state (classical state, angle
between spins 120$^{o}$) and the linear spin-density wave state (LSDW),
containing zig-zag FM chains [${\bf Q}=(\pi,0)$]. For low $U$ there is a
region of no long range order (paramagnetic phase) even at half-filling,
contrary to the square lattice where the band is particle-hole symmetric.

Disorder in this system plays a significant role in determining the
resistivity  
and short-range order, particularly at low temperatures. In fact, the 
resistivity is too high - at 100K the minimum value is about 20$\Omega cm$,
three orders of magnitude larger than the maximum metallic resistivity
for this system (about 10 $m\Omega cm$, assuming a cylindrical FS and 
lattice constants $a,b \simeq 4$\AA\ along the plane). It cannot be only 
disorder that is responsible for this high resistivity in the metallic state,
the interaction surely plays a major role. At the moment we
do not take the disorder into account and look at the clean limit
to investigate the possible magnetic orders that the model provides. 
The effects of disorder will be taken up in a separate 
calculation\cite{TM4} later.

It is also useful to note that the $4f$ levels are not very far from the
Fermi level and a small hybridisation $V_{fd}$ between $4f$ and $5d$
electrons cannot be ruled out completely on the basis of energetics alone
from the band structure results. However, such a term should be very small
owing to reasonably large $U_{ff}$ and at most add a small AF coupling
between the $d$ and $f$ electrons effectively reducing $J_{fd}$ only
slightly.

\begin{figure}[tbp!]
  \begin{center}
\includegraphics[width=7 cm]{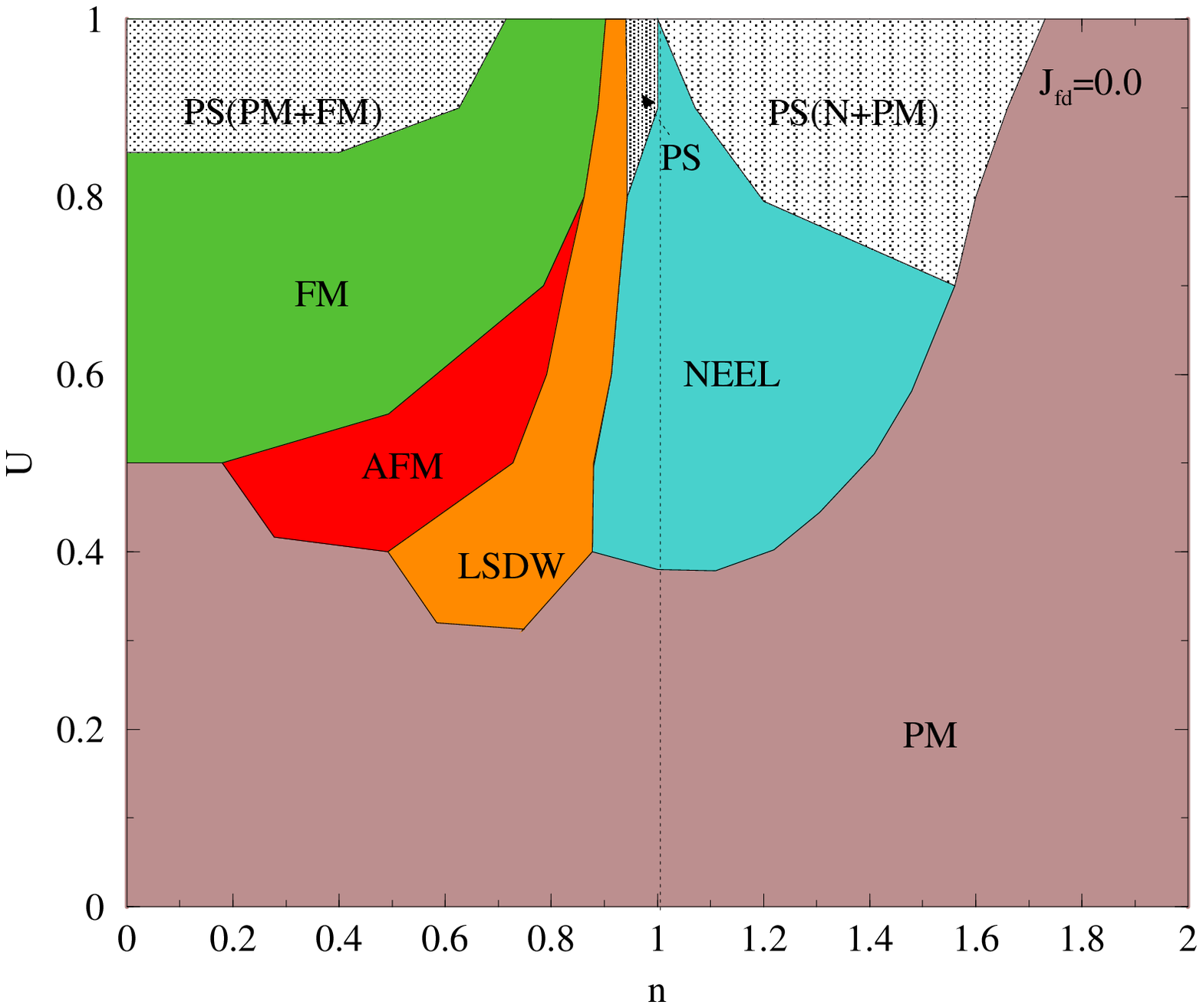}
\includegraphics[width=7 cm]{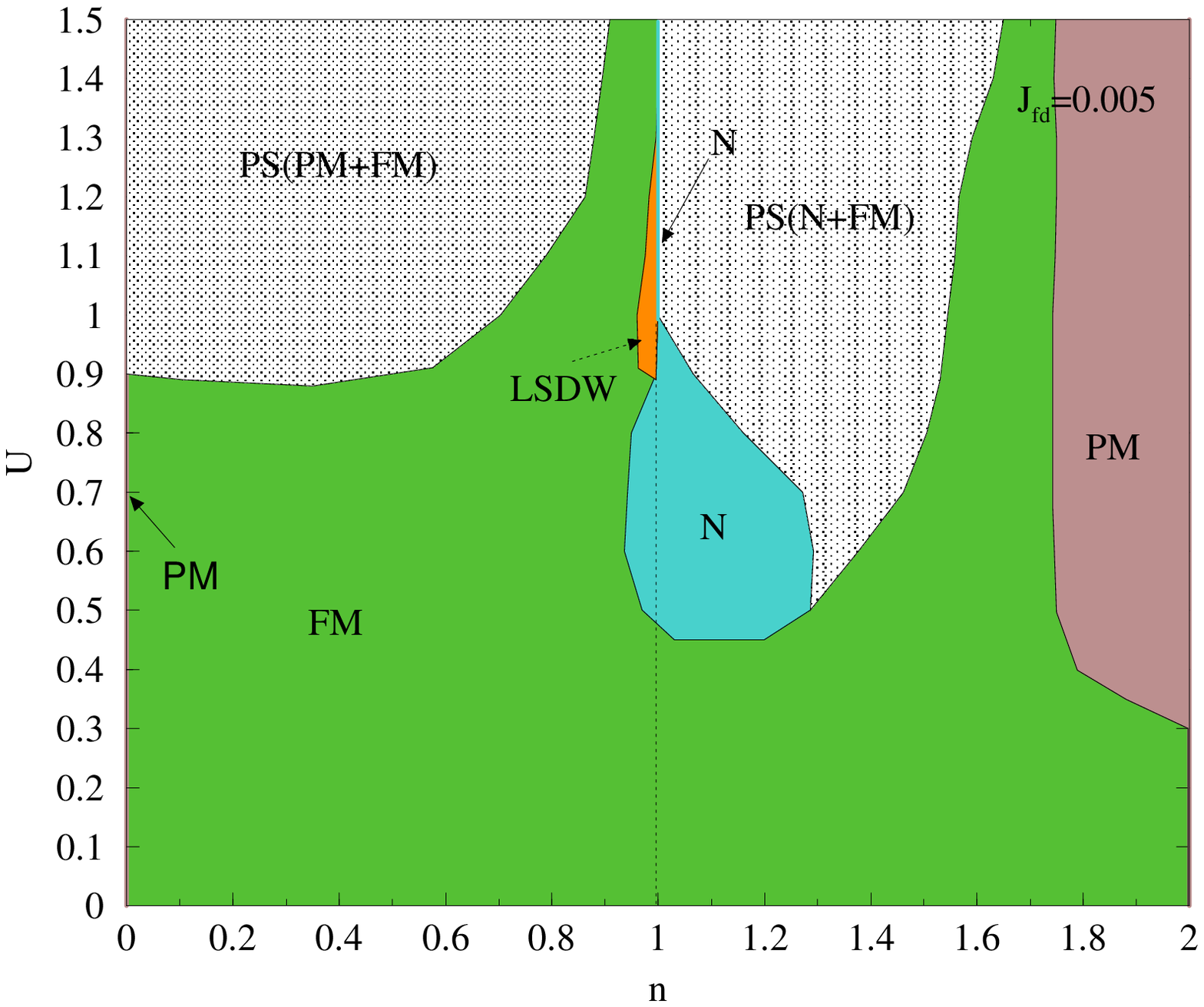}
\includegraphics[width=7 cm]{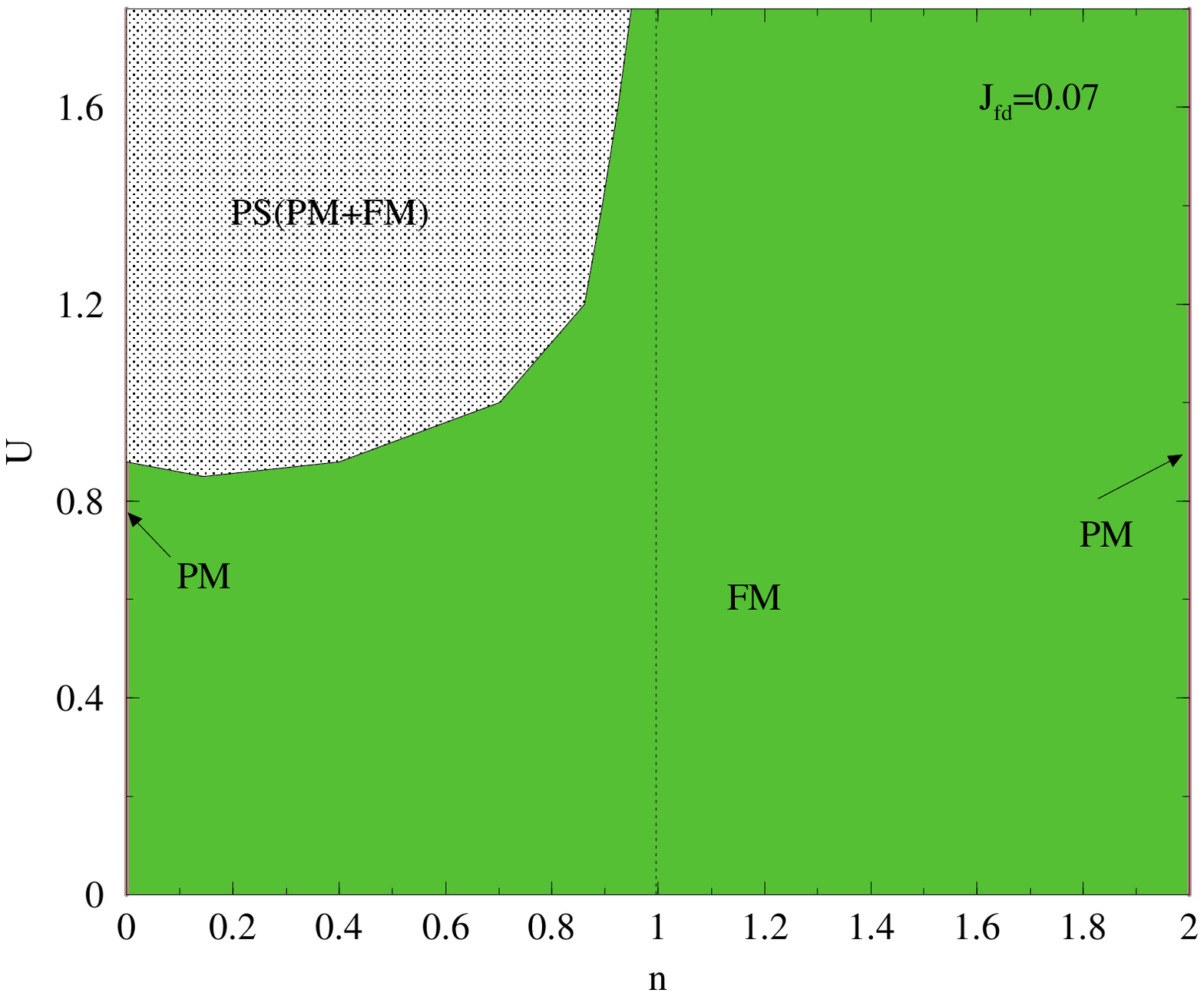}
\end{center}
    \caption{(Color online) Phase diagram in the $U-n$ plane at $J_{fd}=0,0.005,0.07$}
    \label{figure 4}
\end{figure}

\section {calculation and results}

The physics of the model is controlled by competition between the FM 
double exchange and the AF correlations coming from the Coulomb repulsion $U$.
The geometric frustration of the AF state on a triangular lattice adds
to the  number of possible ground states and their closeness in energy.   
In the limit $J_{fd}\rightarrow \infty$, one can project out the direction
of conduction electron spins parallel to the local core spin by using the
transformation 
$e^{i\sigma_{z}\phi_{i}/2}e^{i\sigma_{y}\theta_{i}/2}e^{-i\sigma_{z}\phi_{i}/2}
$. It operates on a two-component spinor and reduces to a system of spinless 
fermions (the angles $\theta_{i}, \phi_{i}$ refer to the polar and azimuthal 
angle of the conduction electron spin with respect to ${\bf S}_i$). 
The on-site Coulomb term, in that case, becomes irrelevant. In this 
situation the AF superexchange term becomes important, though the DE mechanism
would lead to an FM state except close to $n=0$ and $n=2$. 

However, the \gdi and \gdx systems are not really described by this limit
and a finite $J_{fd}$ is the relevant limit here.  We treat 
the core spins classically and the Coulomb interaction term in
a mean-field approximation. To describe the different magnetic states in this 
approximation, we choose ${\bf S=S_{0}}e^{i{\bf Q.r}_i}$, where ${\bf Q}$ 
is the wave vector corresponding to the chosen state described earlier. 
We choose the spin quantization axis in the $xy$ plane. The second term in 
the Hamiltonian Eqn.(1) 
then connects every $k$ point in the Brillouin zone to ${\bf k\pm Q}$ and 
one has to diagonalise the Hamiltonian on a k-grid in the hexagonal BZ. The 
order parameter is obtained by minimising the free energy for a particular 
filling. For the present calculations we have chosen to investigate the 
following ordered phases known from previous investigations on triangular 
lattices. The FM phase, the Ne\`el state, the three-sublattice AFM state
and the LSDW state mentioned earlier. We use the three parameter tight-binding
fit to the band appropriate for \gdi and \gdx systems for our calculations. 

As mentioned above, the Coulomb interactions are generally not very strong
in the $5d$ level, though for the narrow band crossing the Fermi level (see
Fig.1) they may still be very important.  Since both $f-d$ exchange and
$U_{dd}$ contribute to the spin splitting of the conduction band, it is
difficult to specify these two parameters. We therefore allow them to vary
over a possible range, treat them as free parameters and draw the magnetic
phase diagram.
 
In Fig. 4 (a)-(c) the phase diagrams for $J_{fd}=0$, $0.005$ and $0.07 eV$
are shown. The $J_{fd}=0$ phase diagram can be compared with a similar
phase diagram obtained by Hanisch et al.\cite{mh95} with a nn tight-binding
band dispersion $E_{k}=4t \cos {\frac{\sqrt {3}k_{x}}{2}}  \cos
\frac{k_{y}}{2} +2t \cos k_{y}$. It is noticed that there are considerable
differences between the two calculations. At half-filling, $n=1$, we
observe the Ne\`el state for $U$ close to 0.4$eV$ (about $0.36W$, where $W$
is the bandwidth).  This continues all the way upto $U=1eV$. The region of
Ne\`el state is quite wide in our phase diagram around $n=1$. For
$U<0.39eV$, we obtain a paramagnetic phase all the way down. Hanisch et
al. report three-sublattice AF phase at half-filling at
$U_{red}=\frac{U}{U+Uc} = 0.25$ ($U_c=15.81t$ for the triangular lattice,
so that $U=0.58W$) similar to what has been observed
earlier\cite{hrk90}. Away from half-filling we get a small region of phase
segregation (PS) followed by LSDW, AFM, FM and finally another region of PS
depending on the value of $U$ on the hole-doped side, while for $n>1$ there
is a large region of phase separation from the paramagnetic phase.

The phase diagram changes dramatically with finite values of $J_{fd}$.  As
seen in Fig. 4(b) the PM phase reduces to a region close to the empty
($n=0$) and full ($n=2$) bands.  Due to the asymmetry in the DOS, the
regions of stability of the PM phase for $n=0$ and $n=2$ are very
different, the latter being larger. The rest of the PM region becomes FM
for finite $f-d$ exchange as expected. The AFM region completely disappears
while the Ne\`el region becomes narrow beyond $U=1eV$. The LSDW phase also
shrinks considerably. With increasing $J_{fd}$ the FM region takes over
completely (Fig. 4(c)). Two narrow PM regions survive at the ends owing to
negligible KE contributions.

It is obvious from the phase diagrams shown that the value of $J_{fd}$ for
\gdx systems should be very small but finite. In the model without
disorder, the phases are extremely sensitive to changes in the $f-d$
exchange. This is suggestive of the fact that indeed in \gdx several phases
appear very close by in energy. Just as in other similar correlated systems
with competing interactions, e.g., the manganites\cite{dagrev01} and the
double perovskites\cite{atfg04}, a fine tuning of the model parameters is
necessary to specify the phases of the system.  Such a situation leads to
multiple phases with possible microscopic phase segregation as we observe
in Fig. 4. The appearance of a multitude of phase segregated regions in the
present case also has to do with the underlying frustration of the AF
states on a triangular lattice. The large entropy present in the ground
state or very close to it make the ground states easily tunable.

In view of such sensitivity of the phases to $J_{fd}$, we draw the phase
diagram also in $n-J_{fd}$ plane for a fixed value of $U=0.8eV$. Fig. 5
reveals that in order to obtain multiple phases, $J_{fd}$ should be less
than about 17.5$meV$. There are the phases LSDW, Ne\`el, FM and PM present
in the phase diagram with a region of phase separation coming from the
first order transition between the Ne\`el and FM phases. For bigger values
of $J_{fd}$ the entire phase diagram is ferromagnetic. Such a low value for
$f-d$ exchange indicates that the spin splitting in the $d-$band is not due
entirely to the $f-d$ exchange. Instead a considerable part must come from
correlations in the conduction band.

\begin{figure}[tbp!]
  \begin{center}
\includegraphics[width=7 cm]{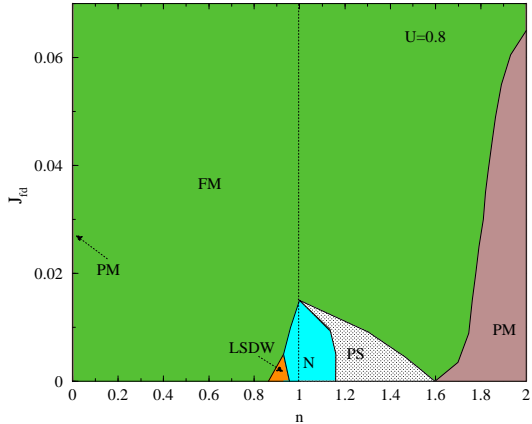}
\end{center}
    \caption{(Color online) Phase diagram in the $J_{fd}-n$ plane for $U=0.8$} 
    \label{figure 5}
\end{figure}

The phase diagram presented in Ryazanov et al.,\cite{rya05} does not seem to
show regions of phase segregation or AF spin order. On the other hand they do 
indicate the presence of a spin glass region (or a mixture of FM and SG) 
beyond $n=0.33$ as can be inferred from the broad susceptibility peak they
observe. Doping 
beyond $n\approx 0.7$ suppresses long range magnetic order completely and 
leads to a broad paramagnetic region.  

The region $n<0.7$ showed large thermal hysteresis in the susceptibility
data. There is a splitting between the field-cooled and zero field-cooled
susceptibility at low temperature indicating disordered magnetic behaviour
(arising from domain rotations, magnetic clusters or superparamagnetic
regions). The ferromagnetic transition temperature is also very sensitive
to the method of sample preparation and the measurement techniques.  As
reported by the authors\cite{private}, the data are quite noisy and the
system appears to be intrinsically disordered, no matter how carefully
prepared. The high value of resistivity in the metallic phase and the
unsaturated magnetic moment deep inside the FM phase are suggestive of the
presence of disorder as well as correlation. All this indicate a possible
competition between different ground states, high degree of magnetic
disorder and a possible microscopic phase segregations. As in the observed
phase diagram, there is a predominance of the FM region close to
half-filling that we also find, but there are other phases close by
depending on the value of $f-d$ exchange coupling in the calculated phase
diagram. The actual value of hole doping in \gdx for a given value of $x$,
is difficult to ascertain\cite{rya05}, though increasing $x$ is understood
to have the effect of hole doping. It is expected from our calculations
that there would be several different phases and microscopic phase
segregations in this system too.

The large magnetoresistance in \gdi has already been attributed
\cite{ETF+01} to the quenching of spin disorder scattering of the
conduction electrons. In a microscopically phase segregated system, there
is likely to be very large magnetic scattering across microscale ordered
regions. In an aligning field such scattering is reduced drastically
leading to a drop in resistivity. In the model, we had a superexchange
interaction to begin with which we have neglected in the calculations that
followed. The presence of such a term would increase AF tendencies and
thereby allow for a larger value of $J_{fd}$ than we derived
above. Besides, the end regions of the phase diagram, $n=0, 2$ will also
become antiferromagnetic owing to the absence of itinerant electrons there.

In conclusion, we have studied a model which, we believe, describes some of
the features of the nearly two dimensional, triangular lattice magnetic
system \gdx.  The resulting phase diagram has some broad similarity with
the experimental one, though it predicts much more structure in the ground
state phases than has so far been observed. We hope to motivate more
experiments in these directions in order to check on our predictions.

\section{acknowledgement}

We are indebted to R. K. Kremer and M. Ryazanov for communicating to us some 
of their results prior to publication and for clarifying some of the experimental 
points involved. 

 

\end{document}